\documentstyle[prd,aps,epsfig]{revtex}
\begin{document}
\title{On vanishing of the factorizable $c\bar c$ contribution 
in the radiative $B \to K^*\gamma$ decay}
\author{Dmitri Melikhov\thanks{Alexander-von-Humboldt fellow. On leave from 
{\it Nuclear Physics Institute, Moscow State University, 119899, Moscow, Russia}}}
\address{ Institut f\"ur Theoretische Physik, Universit\"at Heidelberg, 
Philosophenweg 16,  D-69120, Heidelberg, Germany}
\maketitle

\begin{abstract} 
We argue that the {\it factorizable} $c\bar c$ long- and short-distance contributions 
to the $B\to K^*\gamma$ amplitude vanish, separately, if defined in a gauge-invariant way. 
Therefore, the $c\bar c$ states contribute to the radiative decay only through the 
{\it non-factorizable} soft-gluon exchanges. 
\end{abstract}

\vspace{1cm}

The understanding of the long-distance effects in rare $B$-decays 
is an important theoretical problem. 
In the exclusive radiative decay $B\to K^*\gamma$ there are two types of the 
long-distance contributions: (i) contributions due to the electromagnetic penguin 
bilinear quark operator, and (ii) contributions induced by the four-quark operators 
in the effective hamiltonian. 

The contribution (i) is relatively simple and is described by the $B\to K^*$ form factor.  
The four-quark operators (ii) lead to several contributions of the various types \cite{gp}:  
contributions of the intermediate $c\bar c$ continuum and bound ($\psi, \psi'$, ...) states; 
the weak annihilation; and other more complicated effects. 

The contribution of the $c\bar c$ states is of great importance for the semileptonic 
$B\to (K,K^*)l^+l^-$ decay: the $c\bar c$ vector resonances appear in the physical region 
at $q^2=M_{\rm res}^2$, $q$ the momentum of the lepton pair. There are various models 
\cite{deshpande,kruger,ahmady} describing the $c\bar c$ contribution as function of $q^2$ 
based on the factorization \cite{fact}. 

The $c\bar c$ states also influence the amplitude of the radiative $B\to K^*\gamma$ 
decay which occurs at $q^2=0$. The latter is the only exclusive rare decay observed 
at the experiment, and its reliable theoretical description is strongly needed.  

\noindent The aim of this letter is to point out the following: 
\begin{itemize}
\item[]
The gauge invariance requires that the full {\it factorizable} $c \bar c$ contribution  
(the sum of the long- and short-distance contributions) to the $B\to K^*\gamma$ vanish at $q^2=0$. 
If defined in a gauge-invariant way, the {\it factorizable} long- and short-distance contributions 
vanish, separately. 
\end{itemize}
The $c\bar c$ states in the $q^2$ channel therefore contribute to the amplitude of the 
{\it real} photon emission only through the {\it non-factorizable} soft-gluon exchanges. 
The {\it factorizable} $c\bar c$ contributions are essential for the emission of the 
{\it virtual} photon. 

The amplitude describing the rare radiative decay reads
\begin{eqnarray}
A=\langle K^*\gamma|H_{eff}(b\to s)|B\rangle,  
\end{eqnarray}
where \cite{heff}
\begin{eqnarray}
\label{Heff}
H_{eff}(b\to s) &=& \frac{G_F}{\sqrt{2}}\,\xi_t 
C_{7\gamma}{\cal O}_{7\gamma}
-\frac{G_F}{\sqrt{2}}\,\xi_c 
\left (C_1(\mu){\cal O}_1+C_2(\mu){\cal O}_2\right)
\end{eqnarray}
with $G_F$ the Fermi constant, $\xi_q=V^*_{sq}V_{bq}$, 
$C_i$ the short-distance Wilson coefficients and ${\cal O}_i$ the basis operators 
\begin{eqnarray}
{\cal O}_{7\gamma} &=& \frac{em_b}{8\pi^2}\,
\bar d_{\alpha}\sigma_{\mu\nu}\left (1+\gamma_5\right )\, b_{\alpha}\, F^{(\gamma)}_{\mu\nu}, \\
{\cal O}_1 &=& \bar s_{\alpha}\gamma_{\nu}(1-\gamma_5)c_{\alpha}\;
\bar c_{\beta}\gamma_{\nu}(1-\gamma_5) b_{\beta},\nonumber
\\
{\cal O}_2 &=& \bar s_{\alpha}\gamma_{\nu}(1-\gamma_5) c_{\beta}\; 
\bar c_{\beta}\gamma_{\nu}(1-\gamma_5) b_{\alpha}. 
\end{eqnarray}
The operator ${\cal O}_7$ leads to the penguin form factor $T^{B\to K^*}_1(0)$ known quite well
\cite{ms}. We shall discuss the contribution of the 4-fermion operators ${\cal O}_{1,2}$, i.e. 
the amplitude 
\begin{eqnarray}
A\sim \langle K^*\gamma|\bar s\gamma_\nu(1-\gamma_5)b\;  
\bar c\gamma_\nu(1-\gamma_5)c|B\rangle. 
\end{eqnarray}
Assuming factorization \cite{fact} this ampitude takes the form 
\begin{eqnarray}
\label{fact}
A^{fact}\sim a_2\langle K^*|\bar s\gamma_\nu(1-\gamma_5)b |B\rangle 
      \langle \gamma|\bar c\gamma_\nu(1-\gamma_5)c |0\rangle,  
\end{eqnarray}
where $a_2=C_1+C_2/N_c$. 
The $B\to K^*$ amplitude in this expression is given in terms of the known $B\to K^*$ weak 
form factors. The photon amplitude $\langle \gamma|\bar c\gamma_\nu(1-\gamma_5)c |0\rangle$ 
contains contributions of the $c\bar c$ states, both resonances and continuum. 
There were attempts to model the contribution of the 
$c\bar c$ resonances to the photon amplitude and to estimate in this way the long-distance 
effects in the $B\to K^*\gamma$ decay \cite{gp}. We are going to show however that the 
long-distance contribution to this amplitude vanish (as well as the short-distance one) 
if defined in a gauge-invariant way. 
 
We start our discussion with the photon amplitude from Eq. (\ref{fact}). One finds 
\begin{eqnarray}
\langle \gamma| \bar c\gamma_\nu(1-\gamma_5)c|0\rangle=e\epsilon_\mu(q)\Pi^{c\bar c}_{\mu\nu}(q),
\end{eqnarray}
where 
\begin{eqnarray}
\label{def}
\Pi^{c\bar c}_{\mu\nu}(q)=i\int dx e^{iqx}\langle 0|T(\bar c\gamma_\mu c(x), 
\bar c\gamma_\nu c(0))|0\rangle
\end{eqnarray}
is the charm contribution to the polarization of the vacuum.  
The conservation of the charm vector current $\partial_\mu (\bar c\gamma^\mu c)=0$ leads to the 
transversity of $\Pi^{c\bar c}_{\mu\nu}$ such that it takes the form 
\begin{eqnarray}
\Pi^{c\bar c}_{\mu\nu}(q)=i\left(g_{\mu\nu}-q_\mu q_\nu/q^2\right) \Pi^{c\bar c}(q^2). 
\end{eqnarray}
Let us follow the analysis of the long-distance $c\bar c$ contributions 
of Refs \cite{gp,deshpande}: The function $\Pi^{c\bar c}(q^2)$ contains poles 
at $q^2=M_n^2$, where $M_n$ is the mass of the $c\bar c$ vector resonance 
($\psi_n=\psi,\psi',\dots$), and the contribution of the $c\bar c$ continuum.   
Neglecting the resonance widths we obtain in the region $q^2\simeq M_n^2$  
\begin{eqnarray}
\label{res2}
\Pi^{c\bar c}_{\mu\nu}(q)=-iM_n^2 f^2_n
\left(g_{\mu\nu}-\frac{q_\mu q_\nu}{M_n^2}\right)\frac{1}{M_n^2-q^2}
+\;{\rm regular \;terms}, 
\end{eqnarray}
where $f_n$ is the leptonic decay constant of the vector resonance defined as follows 
\begin{eqnarray}
\label{fv}
\langle 0|\bar c \gamma_\mu c|\psi_n\rangle=\epsilon^{(n)}_\mu f_n M_n. 
\end{eqnarray}
Then one calculates the contribution of the individual resonance to the factorized $B\to K^*\gamma$ 
amplitude of Eq (\ref{fact}), takes a sum over all $c\bar c$ resonances and obtains in this 
way $A^{fact}_{LD}$. The latter is given in terms of the $B\to K^*$ weak transition form factors 
at $q^2=0$ (see definitions in \cite{ms}):  
\begin{eqnarray}
\label{awrong}
A^{fact}_{LD}&=&\sum_n A^{fact}_n(B\to K^*\psi_n\to K^*\gamma)
\nonumber\\
&=&\frac{2}{3}e\frac{G_F}{\sqrt{2}}a_2 \frac{\sum\limits_n f_n^2}{M_B+M_{K^*}}
\epsilon^{*\mu}_{(\gamma)} \epsilon^{*\nu}_{(K^*)}
\left\{i\epsilon_{\mu\nu\alpha\beta}P^\alpha q^\beta
+g_{\mu\nu}A_1(0)(M_B+M_{K^*})^2+P_\mu q_\nu A_2(0)\right\},  
\end{eqnarray}
where $P=p_{B}+p_{K^*}$, $q=p_{B}-p_{K^*}$, $\epsilon$ the polarization vectors.  
Following \cite{gp,deshpande} $A^{fact}_{LD}$ is expected to describe the long-distance 
contribution to the amplitude of the radiative decay. 

Clearly, this amplitude (as well as the contribution of the individual resonance) 
is not gauge-invariant if the form factors do not satisfy the relation
\begin{eqnarray}
\label{ffs}
A_1(0)=-\frac{M_1-M_2}{M_1+M_2}A_2(0). 
\end{eqnarray} 
There were arguments that this relation is approximately satisifed in the leading order of the 
large-energy limit \cite{gp}. This does not help much, since one needs the relation (\ref{ffs}) 
to be exact. But the $B\to K^*$ form factors have no reason to satisfy this relation precisely, 
and therefore both $A^{fact}_n$ and $A^{fact}_{LD}$ are not gauge-invariant.\footnote{
This becomes even more obvious when nonzero $q^2$ are considered: in this case transversity 
of the resonance amplitude requires exact relations between the form factors valid for all 
$q^2$. Clearly, the form factors do not satisfy such relations.}  
This means that the amplitude $A^{fact}_{LD}$ of Eq. (\ref{awrong}) is a gauge-dependent 
quantity and has no clear physical interpretation. In particular, it cannot be used as an 
estimate of the long-distance effects in the radiative decay (cf. \cite{gp}). 

Let us understand better the origin of the difficulty:  

We consider the contribution of the resonance to the gauge-invariant amplitude 
$\Pi^{c\bar c}_{\mu\nu}$. Clearly, this contribution is gauge-invariant near the pole 
where the resonance dominates the amplitude. 

Far from the pole, however, one can describe the individual resonance contribution 
in different ways: it is one of the many regular contributions to the amplitude. 
As one of the possibilities, the contribution of the resonance {\it far} from the pole can 
be defined in a gauge-dependent way - and that is what happened in the example above.  

Obviously, this is allowed: nothing prevents us from splitting the gauge-invariant amplitude 
into many gauge-dependent parts. The only requirement is that the full amplitude - in our 
example the sum of the resonance and continuum $c\bar c $ states - is gauge invariant.  
Working with such gauge-dependent parts is however inconvenient and can lead to a confusion 
in the interpretation of the results.
Much better way is to define the contribution of the 
individual resonance in an explicitly gauge-invariant way. 

%
The direct consequence of the gauge invariance is the relation 
\begin{eqnarray}
\label{polar}
\Pi^{c\bar c}(q^2=0)=0. 
\end{eqnarray} 
This is a well-known property which corresponds to the non-renormalizability of the photon 
mass, and this is an exact relation. Therefore it is convenient to work with the spectral 
representation for the polarization operator which requires a subtraction according to (\ref{polar})
\begin{eqnarray}
\label{pi}
\Pi^{c\bar c}(q^2)=\frac{q^2}{\pi}\int \frac{ds}{(s-p^2)s}\;{\rm Im}\; \Pi^{c\bar c}(s). 
\end{eqnarray} 
The imaginary part contains contributions of the resonances and the continuum states
\begin{eqnarray}
\label{impi}
{\rm Im}\;\Pi^{c\bar c}(s)=\sum_n  f_n^2 \delta(s-M_n^2)+{\rm Im}\;\Pi^{c\bar c}_{cont}(s). 
\end{eqnarray} 
The expressions (\ref{pi}) and (\ref{impi}) for $\Pi^{c\bar c}(q^2)$ lead to an explicitly 
gauge-invariant contribution of the individual resonance to the $B\to K^*\gamma^*(q^2)$ 
amplitude 
\begin{eqnarray}
\label{acorrect}
\bar A^{fact}_n(B\to &&K^*\psi_n\to K^*\gamma^*)
\sim a_2 \left(\frac{f_n}{M_n}\right)^2\frac{1}{M_B+M_{K^*}}
\epsilon^{*\mu}_{(\gamma)} \epsilon^{*\nu}_{(K^*)}\nonumber\\
&&\times\left\{i\epsilon_{\mu\nu\alpha\beta}P^\alpha q^\beta \;V(q^2)q^2
+(g_{\mu\nu}q^2-q_\mu q_\nu)A_1(q^2)(M_B+M_{K^*})^2
+(P_\mu q^2-q_\mu Pq) q_\nu A_2(q^2)\right\}. 
\end{eqnarray}
This expression can be used to describe the resonance contribution for  
any $q^2$.\footnote{Following \cite{amm}, one can multiply the (gauge-invariant) 
resonance contribution in Eq. (\ref{acorrect}) by the phenomenological constant $\kappa$ 
to describe correctly the branching ratio $BR(B\to \psi_n X\to l^+l^-X)=
BR(B\to \psi_n X)BR(\psi_n \to l^+l^-)$.}
Most interesting for us is that $\bar A^{fact}_n=0$ for $q^2=0$. 

So, the explanation looks as follows: 
 
The gauge invariant requires $\Pi^{c\bar c}(0)=0$, and as a result of this relation 
we find for $q^2=0$ 
\begin{eqnarray}
\label{fact3}
\sum_n A^{fact}_n+A^{fact}_{continuum}=0.  
\end{eqnarray}
If we do not take care about the gauge invariance, then each of these 
contributions, separately, is ambiguos and only their sum has the physical interpretation.   
If we define both contributions in a gauge-invariant way as given by (\ref{acorrect}) 
then each of them vanishes for the real photon emission. 

Summarizing, we come to the following conclusion: 
\begin{itemize}
\item
The full {\it factorizable} contribution of the 
$c\bar c$ states to the $B\to K^*\gamma$ amplitude vanishes at $q^2=0$ as the direct 
consequence of the gauge invariance. 
If defined in a gauge-invariant way, the long-distance (resonance) and short-distance 
contributions to the amplitude of the radiative $B\to K^*\gamma$ decay, separately, also vanish. 
\item
Thus, the contribution of the $c\bar c$ states to the amplitude of the {\it radiative} decay 
is completely non-factorizable, as has been already noticed in the literature \cite{krsw,mns}. 
As a consequence, the long-distance contribution cannot be expressed in terms of the $B\to K^*$ 
form factors at $q^2=0$, but requires other relevant quantites for the description of the 
$B\to K^*$ amplitude. For example, in \cite{krsw} the non-factorizable $c\bar c$ contribution 
at $q^2=0$ was expressed in terms of the matrix element of the new - quark-gluon-photon operator. 
\end{itemize}
At $q^2\ne 0$ relevant for the semileptonic $B\to (K,K^*)l^+l^-$ decay, the factorizable 
contributions of the $c\bar c$ states do not vanish. A gauge-invariant modelling of this 
factorizable contribution of the individual $c\bar c$ resonance applicable at any $q^2$ 
was discussed.  

We discussed the $B\to K^*\gamma$ decay, but the same arguments apply to other weak 
radiative $B\to V\gamma$ decays, $V$ the vector meson.  
 

\begin{thebibliography}{30}
\bibitem{gp} B. Grinstein and D. Pirjol, Phys. Rev. D {\bf 62}, 093002 (2000) and refs therein.
\bibitem{deshpande} N. G. Deshpande, J. Trampetic, K. Panose, Phys. Lett. B {\bf 214}, 467 (1988). 
\bibitem{kruger}F. Kr\"uger and L. M. Sehgal, Phys. Lett. B {\bf 380}, 199 (1996). 
\bibitem{ahmady} M. Ahmady, Phys. Rev. D {\bf 53}, 2843 (1996).
\bibitem{fact}M. Neubert and B. Stech, in {\it Heavy flavours II}, A. Buras, M. Lindner (eds),  
World Scientific, Singapore [hep-ph/9705292]. 
\bibitem{heff} B. Grinstein, M. B. Wise and M. J. Savage, Nucl. Phys. {\bf B319}, 271 (1989). 
\bibitem{ms} D. Melikhov, N. Nikitin, S. Simula, Phys. Rev. D {\bf 57}, 6814 (1998); \\ 
D. Melikhov and B. Stech, Phys. Rev. D {\bf 62}, 014006 (2000) and refs therein.  
\bibitem{amm} A. Ali, T. mannel, T. Morozumi, Phys. Lett. B {\bf 273}, 505 (1991). 
\bibitem{krsw}A. Khodjamirian, R. R\"uckl, G. Stoll, D. Wyler, Phys. Lett. B {\bf 402}, 167 (1997). 
\bibitem{mns} D. Melikhov, N. Nikitin, S. Simula, Phys. Lett. B {\bf 430}, 332 (1998).  
\end{thebibliography}
\end{document}